\documentclass[prd,twocolumn,amsmath,amssymb,floatfix,superscriptaddress]{revtex4-1}

\usepackage{graphicx}
\usepackage{amssymb}
\usepackage{amsmath}
\usepackage{color}
\usepackage{ wasysym }
\usepackage{hyperref}
 \usepackage{tabu}

\def\barray{\begin{array}}
\def\earray{\end{array}}
\def\be{\begin{equation}}
\def\ee{\end{equation}}
\def\ben{\begin{equation} \nonumber}
\def\een{\end{equation}}
\def\ban{\begin{eqnarray*}}
\def\ean{\end{eqnarray*}}
\def\ba{\begin{eqnarray}}
\def\ea{\end{eqnarray}}

\def\({\left(}
\def\){\right)}

\def\mnras{MNRAS}

\def\physrep{Phys. Rep.}

\def\apss{Ap\&SS}               

\def\prd{Phys. Rev. D}

\graphicspath{{./fig/}}

\begin{document}

\title{Growth of perturbations in dark energy parametrization scenarios}
\author{Ahmad Mehrabi }

\affiliation{Department of Physics, Bu-Ali Sina University, Hamedan
65178, 016016, Iran}

\begin{abstract}
In this paper we study the evolution of dark matter perturbations in the linear regime by considering the possibility of dark energy perturbations. To do this, two popular parameterizations, CPL and BA with same number of free parameters and different redshift dependency have been considered. We integrate the full relativistic equations to obtain the growth of matter fluctuations for both clustering and smooth versions of CPL and BA dark energy.  The growth rate is larger (smaller) than the $\Lambda$CDM in the smooth cases when $w<-1$ ($w>-1$) but the dark energy clustering gives a larger (smaller) growth index when $w>-1$ ($w<-1$). We measure the relative difference of the growth rate with respect to concordance $\Lambda$CDM and study how it changes depending on the free parameters. Furthermore, it is found that 
the difference of growth rates between smooth CPL and BA is negligible, less than $0.5\%$, while for clustering case, the difference is considerable and might be as large as 2$\%$. Eventually, using the latest geometrical and growth rate observational data, we perform an overall likelihood analysis and show that both smooth and clustering cases of CPL and BA parameterizations are consistent with observations. In particular, we find the dark energy FoM $\sim70$ for the BA and $\sim30$ for the CPL which indicates BA model constrains relatively better than CPL one. 
\end{abstract}
\maketitle

\section{Introduction}
Dark energy (DE) is one of fabulous concepts in modern cosmology introduced to explain the current acceleration expansion of Universe. Several distinct and independent observations including Type I supernovae \citep{Perlmutter:1996ds,Perlmutter:1997zf,Perlmutter:1998np,Riess:2004nr}, the cosmic microwave background (CMB) \citep{Bennett:2003bz,Spergel:2003cb,Spergel:2006hy,Planck:2015xua,Ade:2015yua}, baryon acoustic oscillation (BAO) \citep{Eisenstein:2005su,Seo:2005ys,Blake:2011en} and large scale structures (LSS) \citep{Hawkins:2002sg,Tegmark:2003ud,Cole:2005sx} indicate that the current expansion of the Universe is accelerated.We know that, in the framework of General Relativity (GR), the gravitational force of ordinary matter pushes everything together. So current accelerating expansion of Universe requires an unusual component with negative pressure to overcome the gravity. On the other hand, one can assume the modification of gravity on large scales beyond GR to interpret the cosmic acceleration. The earliest and simplest candidate for DE is cosmological constant ($\Lambda$) which has a negative pressure exactly equal to its energy density ($w_{\Lambda}=\frac{p_{\Lambda}}{\rho_{\Lambda}}=-1$) \citep{Weinberg:1989,Sahni:1999gb,Peebles:2002gy}. Such component has no evolution during the cosmic history ($\rho_{\Lambda}=cte$) and miraculously dominates at recent time (coincidence problem). The cosmological constant acts like a vacuum energy and by considering the cold dark matter (CDM) as another component, one can make a cosmological model the so-called $\Lambda$CDM which is highly consistent with current observations. However the $\Lambda$CDM model suffers to severe theoretical fine-tuning and cosmic coincidence problems (see for example \citep{Sahni2000,Weinberg1989,Carroll2001,Peebles2003,Padmanabhan2003,Copeland2006}).

As we mentioned above, modification of gravity is one solution to explain the current observations. In this approach the laws of  gravity change so that the accelerating expansion of universe is realized without any DE fluid. In this way, the simplest possibility is the modification of Einstein-Hilbert action which is proportional to the scalar curvature (R), and considering a generic function $f(R)$ instead \citep{Schmidt:1990gb,Magnano:1993bd,Dobado:1994qp,Capozziello:2003tk,Carroll:2003wy}. The original form of such models suffers a strong instability \citep{Dolgov:2003px} so people introduce more generalized models to avoid the problem. In addition to $f(R)$ gravity, there are other alternatives including scalar-tensor theories, Gauss-Bonnet gravity and brane-world models which can explain the current observations  \citep{2010deto.book.....A}. Generally, the modification of Einstein gravity leads to additional degrees of freedom and if these come from higher derivatives, the theory suffers from Ostrogradsky ghost instability \citep{Woodard:2006nt}.  

On the other hand, in the framework of GR, we need a fluid with negative pressure to explain current observations. To alleviate the theoretical problems appeared in $\Lambda$CDM theory, we can consider a cosmic fluid with $w\neq-1$. Based on the continuity equation for such a DE fluid, the energy density has an evolution during cosmic history and might alleviate the concordance problem. In addition, the DE fluid can be described by a scalar field in two different approaches: 1- a scalar field with a canonical Lagrangian so-called quintessences models \citep{ArmendarizPicon:2000dh,Copeland:2006wr}. 2- a scalar field with a non-canonical Lagrangian the so-called k-essence models where the negative pressure comes from the kinetic term \citep{ArmendarizPicon:1999rj,ArmendarizPicon:2000dh,Chiba:1999ka,Chiba:2009nh,Amendola:2010}.

DE not only accelerates the cosmic expansion but also affects the evolution of cosmic structures. It is well known that the galaxies and clusters of galaxy that we observe toady are developed from the initial fluctuations at inflation era \citep{Linde1990,Peebles1993}. During the cosmic history, gravity can amplify the amplitude of these fluctuations in particular at the matter dominated epoch. Notice that at DE dominated phase, DE suppresses the fluctuations and slows down the growth rate of structures. Two main properties of DE, that are needed to study in the scenario of cosmological perturbations, are the EoS parameter $w_{\rm de}$ and the effective sound speed $c_e^2=\frac{\delta p}{\delta \rho}$. Notice that at background level the EoS parameter can solely describe the evolution of DE. However, at perturbation level where we study the growth of fluctuations the properties of DE are determined by both EoS and effective sound speed parameters. Two extreme cases have been extensively studied in the literature (see the following text for relevant references) ({\it i}) Models with negligible effective sound speed, $c_e^2\approx 0$. In this case, DE collapses like dark matter (DM) on sub Hubble scales but with much smaller amplitude. ({\it ii}). Models with an effective sound speed roughly equal to unity $c_e^2\approx 1$ ( in unit of light speed $c=1$). In this case DE perturbations can not grow on sub-Hubble scales. 

More deeply speaking, we know that large scale structures (LSS) data provide valuable information regarding the nature of DE \citep{Tegmark:2003ud,Tegmark:2006az}. DE changes the rate of growth and measuring it on large scale structures through redshift space distortion can be used to understand the nature of DE. For scalar based DE models such as quintessence models, the effective sound speed $c_e^2\approx 1$ so DE is smooth on Hubble and smaller scales. On the other hand in the k-essence models, $c_e^2$ can be negligible and so DE perturbations grow through cosmic history \citep{Garriga:1999vw,ArmendarizPicon:1999rj,ArmendarizPicon:2000dh,Babichev:2006vx,Akhoury:2011hr}.     
The possibility of DE clustering and its effects on DM perturbations has been studied in several papers \citep{Bean:2003fb,Hu:2004yd,Ballesteros:2008qk,dePutter:2010vy,Sapone:2012nh,
	Batista:2013oca,Dossett:2013npa,Basse:2013zua,Batista:2014uoa,Pace:2014taa,Steigerwald:2014ava,2015Ap&SS.356..129M,2015PhRvD..92l3513M}. Specifically, authors of \cite{dePutter:2010vy} showed that CMB and LSS slightly prefer a dynamical DE with speed of sound differs from unity. On the other hand, authors of \citep{Basilakos:2009mz} investigated concentration parameter of massive galaxy clusters and pointed out that smooth DE is not consistent well with observations. Furthermore the effects of negligible DE sound speed on growth of DM perturbations in GR framework are studied in \citep{2015MNRAS.452.2930M} and it is revealed that $c_e^2\approx 0$ is favored by observations. 

In this work we focus on the parameterization method to investigate the rule of EoS parameter $w_{\rm de}$ of DE  on the scenario of cosmological structure formation. In the literature, one can find many different EoS parameterizations. On of the simplest and earliest parameterizations introduced by Chevallier-Polarski-Linder is the so-called CPL parameterization \citep{Chevallier2001,Linder2003}.The CPL parameterization is the Taylor expansion of $w_{\rm de}$ with respect to the scale factor $a$ up to first order as $w_{\rm de}(a)=w_{0}+w_{1}(1-a)$ and consequently in terms of redshift as $w_{\rm de}(a)=w_{0}+w_{1}\frac{z}{1+z}$. Notice that although the
CPL is a well-behaved parameterization at early ($z\rightarrow \infty$) and
present ($z=0$) epochs, it diverges at future time ($z=-1$). Beside CPL parameterization some purely phenomenological parameterizations have been proposed ( see \citep{Rezaei:2017yyj} for more details ). The DE clustering scenarios mostly have been studied in wCDM and CPL parameterizations in literature but it is not clear how DE clustering affects DM perturbations in different parameterizations, specifically when the evolution of EoS parameter is different from the CPL. To address this, we investigate the phenomenological parameterization $w(z)=w_0 + w_1\frac{z(1+z)}{1+z^2}$  introduced in \citep{Barboza:2008rh} and hereafter called BA. The BA model provides a different redshift dependency and through it, we can study how effects of DE clustering may change by different parameterizations.
There are many available parameterizations which can be used in current study but we select the BA parametrization for two reasons, firstly it has the same number of free parameters, so it is possible to study the effect of redshift dependency of EoS and secondly it does not diverge at future times.   
 In this paper, we examine how the DE clustering affects the growth rate of DM perturbations in these two parameterizations. Moreover, we study the BA parameterization as a rival model for the CPL in the scenario of cosmological structure formation and use the latest geometrical and growth rate data to examine the ability of these parameterizations against observations.

The structure of this paper is as following: In section
(\ref{sect:back}) we present the basic equations governing the evolution of DE and DM at background and linear perturbation level. In section (\ref{sect:data}), the observational data sets are presented and details of data processing are discussed. Finally in 
(\ref{conclude}) we conclude and discuss our results.

\section{Evolution of DE and DM}\label{sect:back}
In this section we first investigate the evolution of background cosmology and then study the growth of perturbations  considering the CPL and BA DE models. To do these within GR, we need the Einstein field equations along with the continuity equations.  We assume no direct interaction between DM and DE and therefore each fluid evolves independently.

\subsection{Background level}
In the flat FRW Universe the evolution of the Hubble parameter is given by
\begin{equation}\label{eq:scal-fac}
H^2 = H_0^2(\Omega_{dm}(z)+\Omega_{de}(z)+\Omega_{rd}(z)), 
\end{equation} 
where $dm$, $de$ and $rd$ stand for DM, DE and radiation respectively and $\Omega_{x}(z)$ presents the density parameter.  It is convenient to introduce the normalized Hubble parameter $E(z)=\frac{H(z)}{H_0}$, where $H_0$ is the present time Hubble parameter. In the case of non-interacting cosmic fluids, the continuity equation 
leads to $\Omega_{dm}(z)=\Omega_{dm}^{(0)}(1+z)^3$ for DM $(w=0)$, $\Omega_{rd}(z)=\Omega_{rd}^{(0)}(1+z)^4$ for radiation $(w=\frac{1}{3})$ and  
\begin{equation}\label{eq:}
\Omega_{de}(z)=\Omega_{de}^{(0)}(1+z)^3\exp{\int_0^z\frac{w(z)}{1+z}dz} 
\end{equation}
for a DE fluid with an arbitrary EoS parameter. The superscript $(0)$ indicates the present time value of quantities. Notice that in a flat universe $\Omega_{rd}^{(0)}+\Omega_{dm}^{(0)}+\Omega_{de}^{(0)}=1$ and hereafter we  confine ourself to a flat Universe.  
Now we calculate the energy density of DE for two CPL and BA parameterizations considered in this work. The EoS parameter of CPL in terms of redshift  is given by    
\begin{equation}\label{eq:cpl-w}
w_{\rm CPL}(z)=w_0 + w_1\frac{z}{1+z},
\end{equation}
where $w_0$ and $w_1$ are two free parameters of the model. Using continuity equation, the density parameter in this case is 
\begin{equation}\label{eq:cpl-rho}
\Omega_{de}(z) =\Omega^{(0)}_{de}(1+z)^{3(1+w_0+w_1)}\exp{\frac{-3w_1z}{1+z}}. 
\end{equation}
The EoS parameter of BA reads  
\begin{equation}\label{eq:ba-w}
w_{\rm BA}(z)=w_0 + w_1\frac{z(1+z)}{1+z^2},
\end{equation}
with two free parameters like the CPL model. Here the energy density of DE can be easily obtained. In this case the density parameter is
\begin{equation}\label{eq:ba-rho}
\Omega_{de}(z) =\Omega_{de}^{(0)}(1+z)^{3(1+w_0)}(1+z^2)^{\frac{3}{2}w_1}. 
\end{equation}

\begin{figure}[h]
	\centering
	\includegraphics[width=0.5\textwidth]{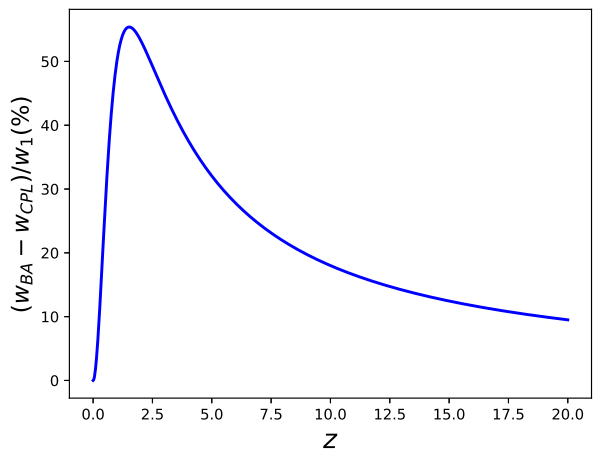}
	\caption{The relative difference of EoS of our models in unite of $w_1$ with respect to the cosmic redshift. }
	\label{fig:delw}
\end{figure}

\begin{figure}[h]
	\centering
	\includegraphics[width=0.5\textwidth]{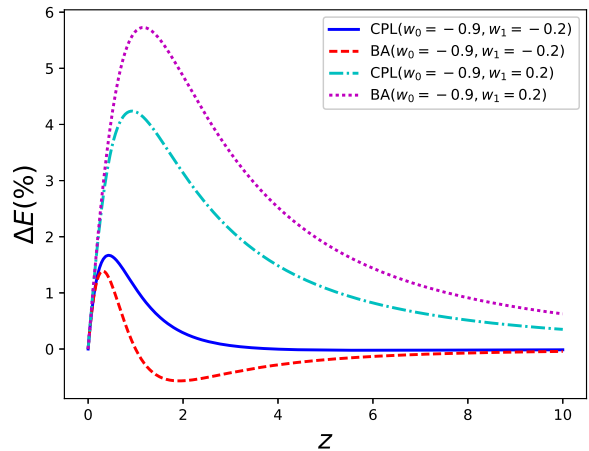}
	\caption{The redshift evolution of percentage relative difference between Hubble parameters of CPL and BA parameterizations with respect to standard $\Lambda$CDM model. }
	\label{fig:rel-e}
\end{figure}

At present and early times these two parameterizations are the same but at far future $z\rightarrow -1$, the CPL diverges while the BA gives a constant value.
These two models differ from each other due to different redshift dependency. We show the difference between two parameterizations, $w_{\rm BA}-w_{\rm CPL}$, in unite of $w_1$ in Fig. (\ref{fig:delw}) which indicates a maximum difference of around $0.55$ at redshift $z\sim 2$. Such difference in the EoS, not only affects the Hubble parameter but also the growth of rate of perturbations. To realize how various redshift dependencies affect the Hubble parameter as well as growth of perturbations, we measure the relative difference of these two quantities in this and subsequent parts.
The relative difference between Hubble parameter with respect to concordance $\Lambda$CDM one is computed as following
\begin{equation}\label{eq:rel-e}
\Delta E(\%) = \frac{E_{\rm CPL,BA}-E_{\Lambda \rm CDM}}{E_{\Lambda \rm CDM}}\times 100,
\end{equation} 
where $E_{\Lambda}$ is the Hubble parameter for the $\Lambda$CDM.
In Fig (\ref{fig:rel-e}) we show the evolution of the relative difference $\Delta E$ as a function of redshift $z$ for different values of free parameters $w_0$ and $w_1$. Here for all cases we fix $w_0$ to $-0.9$ and allow $w_1$ gets $-0.2$ and $+0.2$, only to show how these two models affects the Hubble parameter.

As expected, at present time all cases are coincide to each others because of normalization of the Hubble parameter. We observe that the difference between different parameterizations occurs at low redshifts between $z \sim 1-2$. This result is so interesting since at low redshifts DE dominates the total energy of Universe and the dynamics of the whole Universe is determined by DE. In the case of $w_1=-0.2$, we observe that for both CPL and BA the quantity $\Delta E$ is roughly  $1\%$ at redshift around $z\sim1.7$. While in the case of $w_1=0.2$, this value is approximately  $5-6\%$ at $z\sim1.6$. We also observe that in the case of $w_1=0.2$, the maximum of $\Delta E$ for BA parameterization is roughly $2\%$ larger than CPL one. Moreover for $w_1=-0.2$ ($w_1=0.2$) the BA parameterization results smaller (larger) $\Delta E$ compare to CPL.  At high redshifts, we see that the difference goes to zero for all parameters. This means that at early times ( DM dominated epoch) the effect of DE on the dynamics of Universe is negligible. The Hubble parameter not only affects the evolution of cosmic fluid but also has a direct influence on the evolution of fluctuations. We will see in the next subsection that how the Hubble parameter difference alters the growth of fluctuations.


\subsection{Perturbation level}
In the framework of GR, the evolution of perturbations can be described using
the conformal Newtonian gauge. In this gauge the perturbed FRW metric is given by:
\begin{equation}\label{eq:frw-per}
ds^2 = a(t)^2[(1+2\psi)d\eta^2-(1-2\phi)\delta_{ij}dx^idx^j],
\end{equation}
where $\psi$ and $\phi$ are the Bardeen potentials and $\eta$ is the conformal time. For a fluid without anisotropic stress the Einstein equations imply $\psi=\phi$ but for modify gravity models this is not true generally. Following, we assume the DE fluid has no anisotropic stress so these two potentials are the same.  
The Einstein equations in perturbed FRW metric read
\begin{eqnarray}
3\mathcal{H}\phi^{\prime}&+&\left(3\mathcal{H}^2+k^2\right)\phi =  
-\frac{3\mathcal{H}^2}{2}(\Omega_{\rm m}\delta_{\rm m}+\Omega_{\rm d}\delta_{\rm d})\;,
\label{eq:first-order-En-eq1}\\
\phi^{\prime\prime}&+&3\mathcal{H}\phi^{\prime}+
\left(\frac{2a^{\prime \prime}}{a}-\mathcal{H}^2\right)\phi  = 
\frac{3\mathcal{H}^2}{2}\Omega_{\rm d}\frac{\delta p_{\rm d}}{\delta\rho_{\rm d}}\delta_{\rm d}\;,
\label{eq:first-order-En-eq2}
\end{eqnarray}
where $\mathcal{H}=aH$ is the conformal Hubble parameter and prime denotes derivative with respect to the conformal time. Here $\delta_m$ and $\delta_{d}$ are the perturbations of DM and DE respectively. Notice that in Eqs.( \ref{eq:first-order-En-eq1} \& \ref{eq:first-order-En-eq2}), we consider a general case in which both DM and DE have perturbed. For sub-horizon scales $\mathcal{H}^2\ll k^2$ and in matter domination epoch $\phi\approx cte$, the first equation turns to the usual Poisson equation,   
\begin{equation}\label{eq:pos}
k^2\phi = -\frac{3\mathcal{H}^2}{2}(\Omega_{\rm m}\delta_{\rm m}+\Omega_{\rm d}\delta_{\rm d}).
\end{equation}
The continuity equations at perturbation level for a general fluid are \citep{Ma:1995ey}
\begin{eqnarray}
\delta^{\prime}_{i} & = & -(1+w_i)(\theta_i-3\phi^{\prime})-
3\frac{a^{\prime}}{a}\left(\frac{\delta p_i}{\delta\rho_i}-w_i\right)\delta_i\;,\label{eq:first-order-conser1}\\
\theta^{\prime}_{i} & = & -\frac{a^{\prime}}{a}(1-3w_i)\theta_i-\frac{w_i^{\prime}}{1+w_i}\theta_i+
\frac{\frac{\delta p_i}{\delta \rho_i}}{1+w_i}k^2\delta+k^2\phi\;,\label{eq:first-order-conser2}
\end{eqnarray}
where $\theta$ is divergence of velocity. The ratio of pressure to density perturbation
$\frac{\delta p_i}{\delta \rho_i}$ needs to be a gauge invariant quantity, so it is given by \citep{Bean:2003fb}  
\begin{equation}\label{eq:c_eff}
\frac{\delta p}{\delta\rho}=c_{\rm e}^2+3\mathcal{H}(1+w)(c_{\rm e}^2-c_{\rm a}^2)
\frac{\theta}{\delta}\frac{1}{k^2}\;,
\end{equation}
where $c_e^2$ and $c_a^2$ indicate effective and adiabatic sound speed square respectively. The adiabatic sound speed square is given by:
\begin{equation}\label{eq:c_a}
c_{\rm a}^2=w-\frac{a\frac{dw}{da}}{3(1+w)}\;,
\end{equation}
which is determined by the EoS parameter and so is negative for most of DE models. The negative value of sound speed square ($c_{\rm a}^2<0$) leads to unstable exponentially growth of perturbations. Fortunately, the presure perturbation is given in term of the effective sound speed and not the adiabatic one and the problem can be avoided when we deal with the effective sound speed. In contrast to the adiabatic sound speed, the effective sound speed square is a positive value in the range of $[0,1]$ (for more details about DE sound speed see the next subsection). Here we assume both smooth  ($c_{e} \sim 1$) and clustered ($c_{e}\sim 0$) DE scenarios and study their effects on the growth of matter perturbations within the framework of DE parameterizations considered in this work.

Taking another derivative of Eq.(\ref{eq:first-order-conser1}) and using Eq.( \ref{eq:first-order-En-eq1} \& \ref{eq:first-order-conser2} ), we can obtain a second order differential equation that governs evolution of perturbations for the fluids \citep{2015MNRAS.452.2930M}. These differential equations are in the following forms 
\begin{eqnarray}
\label{eq:sec-ord-delta_m}\frac{d^{2}\delta_{\rm m}}{da^{2}}&+& \frac{1}{a}(2+\frac{\mathcal{H}^{\prime}}{\mathcal{H}^2})\frac{d\delta_{\rm m}}{da}= S\;, \\\label{eq:sec-ord-delta_d}
\frac{d^{2}\delta_{\rm d}}{da^{2}}&+& \frac{1}{a}\left[2+
\frac{\mathcal{H}^{\prime}}{\mathcal{H}^2} +3c_{\rm a}^2-6w_{\rm d}\right]\frac{d\delta_{\rm d}}{da}\\ \nonumber& + &B_d  
\delta_{\rm d}  = (1+w_{\rm d}) S\;, 
\end{eqnarray}
where $B_d$ and $S$ are given by
\begin{eqnarray}
B_{\rm d} & = & \frac{1}{a^2}\left[3\left(c_{\rm e}^2-w_{\rm d}\right)
\left(1+\frac{\mathcal{H}^{\prime}}{\mathcal{H}^2}-
3w_{\rm d}+3c_{\rm a}^2-3c_{\rm e}^2\right)\right. \label{eq:bd} \\  \nonumber      
& + & \left.   \frac{k^2 }{\mathcal{H}^2}c_{\rm e}^2-3a\frac{dw_{\rm d}}{da}\right],\; \\  \nonumber
S & = & \frac{3\mathcal{H}^2}{2}(\Omega_{\rm m}\delta_{\rm m}+\Omega_{\rm d}\delta_{\rm d}).\; \label{eq:s} 
\end{eqnarray}
To obtain the DE perturbation equation we set $c_e^2=0$ and substitute Eq. (\ref{eq:c_eff}) into Eqs.(\ref{eq:first-order-conser1} \& \ref{eq:first-order-conser2}) but in the case of smooth DE, we set $\delta_d=0$ in these equations. Notice that the source term in Eq.(\ref{eq:sec-ord-delta_d}) is proportional to $1+w_{\rm d}$ so any DE perturbation vanishes in the case of standard $\Lambda$CDM cosmology. In Eqs. (\ref{eq:sec-ord-delta_m} \& \ref{eq:sec-ord-delta_d}) we need to know the derivative of the Hubble parameter which can be easily obtained from the Friedman equations as follows
\begin{equation}\label{hdot-over-h2}
\frac{\mathcal{H}^{\prime}}{\mathcal{H}^2}
=-\frac{1}{2}(1+3\Omega_{\rm d}w_{\rm d})\;,
\end{equation}  

In order to solve the perturbation equations (\ref{eq:sec-ord-delta_m} \& \ref{eq:sec-ord-delta_d}), we use the following 
initial conditions \citep{Abramo:2008ip,2015MNRAS.452.2930M}, 
\begin{eqnarray}\label{eq:ini}
\delta_{\rm m,i}&=&-2\phi_{\rm i}\left(1+\frac{k^2}{3\mathcal{H_{\rm i}}^2}\right)\;, \\ 
\frac{d\delta_{\rm m,i}}{da}&=&-\frac{2}{3}\frac{k^2}{\mathcal{H_{\rm i}}^2}\phi_{\rm i}\;,\\
\delta_{\rm d,i}&=&(1+w_{\rm d})\delta_{\rm m,i}\;,\\
\frac{d\delta_{\rm d,i}}{da}&=&(1+w_{\rm d})\frac{d\delta_{\rm m,i}}{da}+\frac{dw_{\rm d}}{da}\delta_{\rm m,i}\;, 
\end{eqnarray}
where we set $\phi_i=-6\times 10^{-7}$. These initial conditions lead to a linear perturbation ($\delta_m\approx 0.1$) for scale $k=0.15h$Mpc$^{-1}$. In this work we fix the scale to $k=0.15h$Mpc$^{-1}$ and integrate the perturbation equations numerically from $a=0.01$ to present time. We should note that varying the value of $k$ to other linear scales has a very tiny effect on the evolution of perturbations, as discussed in \citep{2015MNRAS.452.2930M}. Furthermore, notice that the initial conditions for DE perturbations are given by assuming an adiabatic condition \citep{Kodama:1984,Amendola:2010}.  

To compare our results with observation, the relevant quantity  is the product of growth rate $f$ and mass variance in a sphere of radius 8Mpc/h ($\sigma_8$). The growth rate at any redshift is given by:
\begin{equation}\label{eq:growth-rate}
f(z)=-\frac{1+z}{\delta_m(z)}\frac{d\delta_m(z)}{dz}.
\end{equation} 
Also the mass variance at a given redshift is $\sigma_8(z)=\sigma_{8}\frac{\delta_m(z)}{\delta_m(z=0)}$, where $\sigma_{8}$ is the mass variance at present time. Generally, the mass variance at present time is a free parameter and can be constrained using the observational data.


To realize how DE perturbations affect the growth of DM perturbations, we fix $\Omega_m=0.3$, $h=0.7$ and compute  the relative difference of $f\sigma_8(z)$ for CPL and BA parameterizations with that of in the $\Lambda$CDM according the following relation
\begin{equation}\label{eq:del-fs8}
\Delta f\sigma_8=\frac{f\sigma_{\rm 8,model}-f\sigma_{\rm 8,\Lambda}}{f\sigma_{\rm 8,\Lambda}}\times 100
\end{equation}

\begin{figure}[h]
	\centering
	\includegraphics[width=0.5\textwidth]{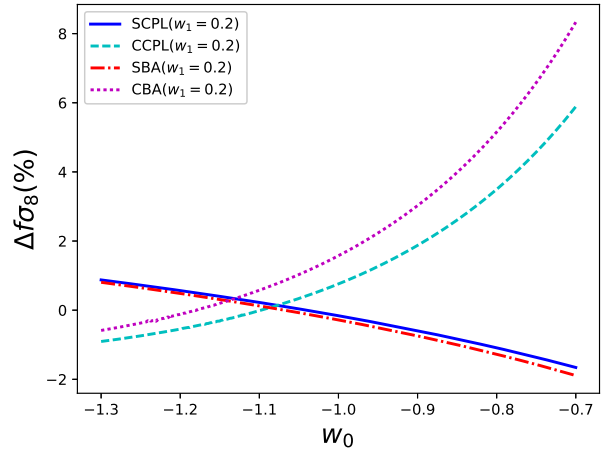}
	\caption{The relative growth rate as a function of $w_0$ for the CPL and BA parameterizations. SCPL and SBA (CCPL and CBA) indicate smooth CPL and BA (clustered CPL and BA) models. }
	\label{fig:delfs81}
\end{figure}

Since our parameterizations have two free parameters, we fix one parameter and present $\Delta f\sigma_8$ as a function of other parameter. In Fig.(\ref{fig:delfs81}) we set $w_1=0.2$ and show the evolution of $\Delta f\sigma_8$ calculated at present time as a function of $w_0$ for both smooth and clustering DE scenarios. The results for smooth CPL and BA are very close to each other and the differences are less than $0.5\%$. However for clustered CPL and BA cases, the difference is relatively large and we measure  it as $0.5-2.5\%$ for $w_0$ in the range $(-1.3,-0.7)$.

Moreover, the smooth and clustered DE behave differently at both side of $w_{\rm d}=-1$. We observe that $\Delta f\sigma_8$ is positive for clustering cases when $w_{\rm d}>-1$ and may reach to $8\%$ for $w_0=-0.7$. On the other hand the relative growth rate is negative for smooth DE parameterizations when $w_{\rm d}>-1$ but the differences with respect to the $\Lambda$CDM model is around $2\%$ for $w_0=-0.7$. In addition, one can measure the small difference between smooth and clustering cases when the EoS parameter crosses the phantom line $(w_{\rm d}<-1)$.  
The different behaviors of clustered DE models at both sides of phantom line $(w_{\rm d}=-1)$ can be described as following. When the EoS of DE crosses the phantom line, the sign of source term in Eq.(\ref{eq:sec-ord-delta_d}) changes and DE perturbations are negative and vise versa.     

Another interesting point that might be realized from Fig. (\ref{fig:delfs81}) is that for DE in regime $w_{\rm d}>-1$ ($w_{\rm d}<-1$), the growth rate of DM perturbations is larger (smaller) compare to the $\Lambda$CDM.  This prediction can be easily understood from an extra term due to DE perturbations ($\delta_d$) in the source term of DM perturbation equation which is positive (negative) in the case $w_{\rm d}>-1$ ($w_{\rm d}<-1$).  
For smooth DE models, there is no $\delta_d$ and the growth of DM perturbations is affected by the evolution of the Hubble parameter. As it is clear from  Fig.\ref{fig:rel-e}, the Hubble quantity is larger than the $\Lambda$CDM for $w_{\rm d}>-1$ regime so the growth rate is smaller.      
\begin{figure}[h]
	\centering
	\includegraphics[width=0.5\textwidth]{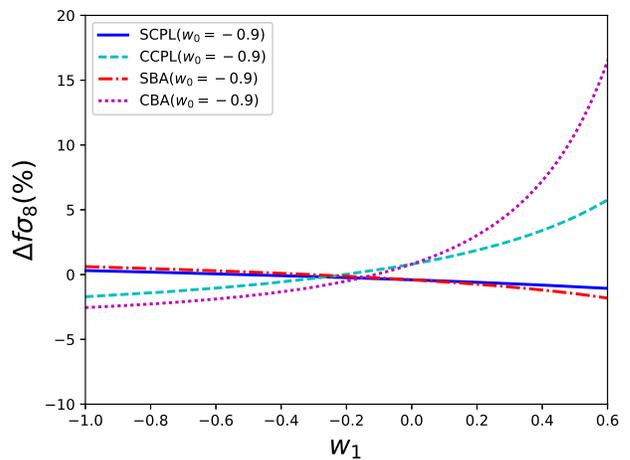}
	\caption{The relative growth rate as a function of $w_1$ for the CPL and BA parameterizations. SCPL and SBA (CCPL and CBA) indicate smooth CPL and BA (clustered CPL and BA) models.}
	\label{fig:delfs82}
\end{figure}

In Fig.(\ref{fig:delfs82}), we show the evolution of  $\Delta f\sigma_8(z=0)$ as a function of $w_1$ where $w_0$ is fixed to $-0.9$. We observe that the evolution  of $\Delta f\sigma_8(z=0)$ with respect to $w_1$ is similar to previous one. However our result shows very small difference for smooth DE models (less than $0.5\%$) when $w_1$ is in range $(-1,0.6)$. For clusterd DE models, when $w_1<0$, the difference is small but for positive value of $w_1$ the difference is relatively larger and might be as large as $10\%$ for $w_1=0.6$. Furthermore, the result shows a tiny difference between our models and the $\Lambda$CDM for $w_1<0$ in both cases of smooth and clustering DE. For $w_1>0$ the smooth DE does not change the growth rate significantly compare to the $\Lambda$CDM, but for clustering case the difference increases rapidly as $w_1$ increases and might be as large as $17\%$ for $w_1=0.6$.       

From the above statements it has been revealed that how EoS parameter of DE affects the DM growth rate in both smooth and clustering DE cases. For smooth DE, these two parameterizations roughly give similar results but in the latter case, the growth rate might considerably be different in these two models which is due to the different redshift dependency.
  Based on our analysis, various redshift dependencies of EoS might change the growth rate around $2\%$ in clustering DE scenarios, which is 4 times larger than the growth rate difference in the smooth DE. Using current observational data, it is not possible to distinguish between smooth and clustering DE cases, but according to \citep{2011PhRvD..84h3523S,Asaba:2013mxj}, a combination of weak lensing and the peculiar velocity observations can break the degeneracy of the DE clustering (no anisotropic stress) and some modify gravity theories and also distinguish between the smooth or clustering DE cases. So further cosmological data, for example data from \emph{Euclid} can improve the quality of data and it might be possible to distinguish between clustering or smooth DE.  
  

\subsection{DE sound speed}
Since the DE sound speed is a crucial quantity in the clustering DE, it is worth to discuss it with more details. In a general case the pressure not only depends on the energy density but also on the entropy $s$ so $p(\rho,s)$ and its perturbation is
\begin{equation}\label{eq:delp}
\delta p = (\frac{\partial p}{\partial \rho})_s \delta \rho +(\frac{\partial p}{\partial s})_{\rho}\delta s\;,
\end{equation}  
where $(\frac{\partial p}{\partial \rho})_s$ is the adiabatic sound speed square and the second term is due to entropy perturbation \citep{2005pfc..book.....M}. Hence we have 

\begin{equation}\label{eq:delp2}
\frac{\delta p}{\delta\rho} = c_a^2  +(\frac{\partial p}{\partial s})_{\rho}\frac{\delta s}{\delta\rho}\;.
\end{equation}
For a perfect fluid there is no entropy perturbation and the second term vanishes. In this case the pressure perturbation is given in term of the adiabatic sound speed square and it is negative for most DE models. However the second term can compensate the first and the pressure perturbation become zero or positive. The sum of these two terms results the effective sound speed of the DE which is the related quantity in the case of DE perturbations. 

As we mentioned, it is generally believed that if the sound speed square for a fluid be negative, its perturbations are unstable. However by considering the perturbation of entropy, the problem can be avoided. The second term in Eq.(\ref{eq:delp}) can be dominated due to some dissipative process and consequently it would change the effective sound speed of DE to a null or positive value \citep{2005pfc..book.....M}. Notice that the above discussion is important when we consider the clustering DE models. In the cases of smooth DE scenarios we need only the EoS parameter of DE to determine the evolution of DM perturbations.

\section{Observational data and likelihood analysis}\label{sect:data}
In order to check the validity of our models with observational data, we perform a MCMC analysis using most recent data. Basically the observational data consist of two parts, 1- data to constrain the background parameters and 2- data to constrain the growth rate of DM perturbations (at the first level). We use the most recent SN Ia (JLA sample), BAO, CMB and the Hubble parameter data to constrain the background parameters including $(\Omega_m,h,w_0,w_1)$  and also the growth rate of perturbations, $f\sigma_8$ data, to constrain our models at first perturbation level. Following, we first briefly explain the data set and then procedure of MCMC analysis. Finally we present the best value of parameters as well as their uncertainties and discuss the results. 

For the JLA SN sample, the theoretical value of distance module $\mu_{th}$ is given by:
\begin{equation}\label{eq:mu-th}
\mu_{th} =  5 \log_{10}(\frac{d_L(z_{hel},z_{cmb})}{\rm {Mpc}}) + 25,
\end{equation}
where $d_L$ is the luminosity distance and $z_{cmb}$ ($z_{hel}$ ) is the CMB rest-frame (heliocentric) redshift of SN. The luminosity distance $d_L$ is given by \citep{Wang:2016bba}
\begin{equation}\label{eq:dl}
d_{L}=(1+z_{hel})r(z_{cmb}),
\end{equation} 
where $r(z)$ is the comoving distance which is given in term of the normalized Hubble parameter. The observational distance module is given by the following empirical relation \citep{Betoule:2014frx}
\begin{equation}\label{eq:mu-obs}
\mu_{obs} = m_{B}- M_{B}+\alpha\times\mathbf{x_1}-\beta\times\mathbf{C},
\end{equation}
where $m_B$ corresponds to the observed peak magnitude in rest frame of B band and $\alpha$, $\beta$ and $M_B$ are nuisance parameters which should be marginalized at end. To see details and definition of other parameters see \citep{Betoule:2014frx}. 
The $\chi^2$ of SN data is given by
\begin{equation}\label{eq:chi2-sn}
\chi^2_{sn} = \Delta\mu^{T}\mathbf{Cov}_{sn}^{-1}\Delta\mu, 
\end{equation}
where $\Delta\mu=\mu_{obs}-\mu_{th}$ and $\mathbf{Cov}_{sn}$ is the total covariance matrix which includes statistical and systematic uncertainties (for more details of covariance matrix see \citep{Betoule:2014frx}).

The next data set is the BAO which is based on the observed baryon oscillations in the power spectrum of galaxy correlation function. In our analysis we use 6 distinct data points which are presented in Tab.(\ref{tab:bao}).
\begin{table}
	\centering
	\caption{The BAO data used in this work.}
	\begin{tabular}{|c|c|c|}
		\hline
		$z$     & $d_i$    & Survey \& References  \\ \hline
		$0.106$ & $0.336$  & 6dF \citep{Beutler:2011hx}\\ \hline
		$0.35$   & $0.113$ & SDSS-DR7 \citep{Padmanabhan:2012hf}\\ 
		$0.57$  & $0.073$ & SDSS-DR9 \citep{Anderson:2012sa} \\ \hline
		$0.44$  & $0.0916$ & WiggleZ \citep{Blake:2011en} \\ 
		$0.6$   & $0.0726$ & WiggleZ \citep{Blake:2011en} \\ 
		$0.73$  & $0.0592$ & WiggleZ \citep{Blake:2011en} \\ \hline 
	\end{tabular}
		\label{tab:bao}
\end{table} 
In this case the quantity $\chi^2_{bao}$ in term of covariance matrix is given by
\begin{equation}\label{eq:xi2-bao}
\chi^2_{\rm bao}=\mathbf{Y}^{T}\mathbf{C}_{\rm bao}^{-1}\mathbf{Y}\;,
\end{equation}
where we use $\mathbf{Y}$ and  $\mathbf{C}_{\rm bao}$ from\cite{Hinshaw:2012aka}.

Since the position of the CMB acoustic peaks depends on the DE dynamic through the angular diameter distance, the CMB data provide valuable information to  constrain a DE model. The process of calculating $\chi^2_{cmb}$ (for Planck data) does not repeat here and we refer reader to \citep{2015MNRAS.452.2930M} for more details. 

Furthermore, we use an updated version of the Hubble parameter compare to our previous one in \citep{2015MNRAS.452.2930M}. The Hubble parameter in this work are those data (38 data points) collected in \citep{Farooq:2016zwm}. For these data set the $\chi^2$ is given by 
\begin{equation}\label{eq:xi2-H}
\chi^2_{\rm h}=\sum_i\frac{[H(z_i)-H_{\rm ob,i}]^2}{\sigma_i^2}\;,
\end{equation} 
where $H(z_i)$ $(H_{\rm ob,i})$ is the theoretical ( observational) Hubble parameter.

In addition to above data,  we use the growth rate data ($f\sigma_8$) which are obtained from RSD. Since not all of current available data points are independent, we use data introduced in  \citep{Nesseris:2017vor} which are independent set of growth rate data.  

\begin{figure}
  	\centering
  	\includegraphics[width=.5\textwidth]{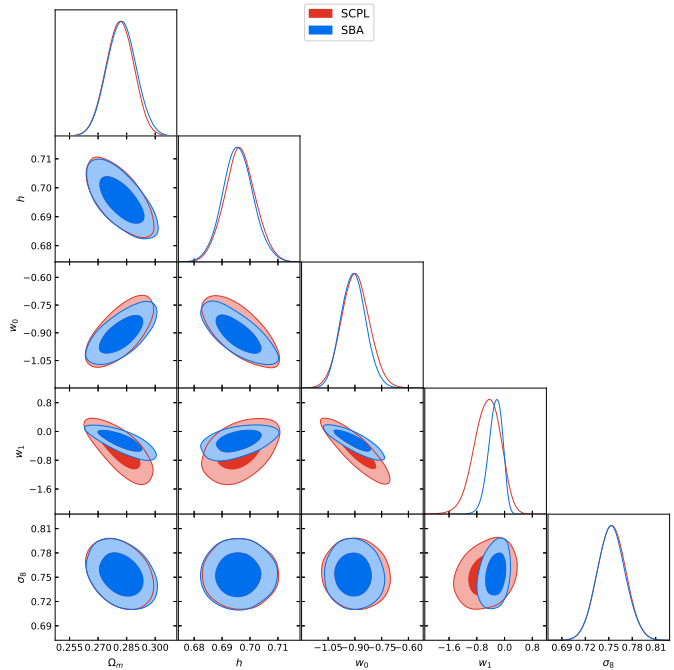}
  	\caption{The confidence regions for smooth DE.}
  	\label{fig:res-sm}
  \end{figure}

\begin{figure}
  	\centering
  	\includegraphics[width=.5\textwidth]{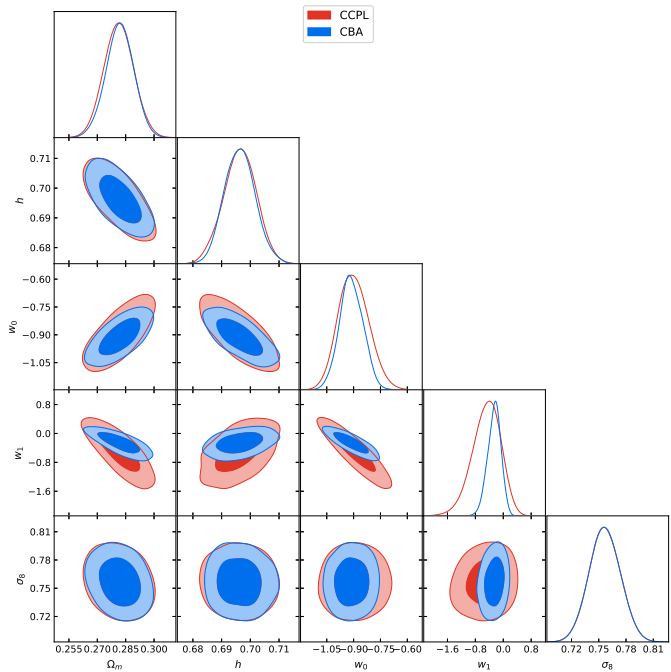}
  	\caption{The confidence regions for clustering DE.}
  	\label{fig:res-cl}
  \end{figure}

Finally, since the overall likelihood is the product of each likelihood, the total $\chi^2$ is given by
\begin{equation}\label{eq:like-tot_chi}
\chi^2_{\rm tot}=\chi^2_{\rm sn}+\chi^2_{\rm bao}+\chi^2_{\rm cmb}+\chi^2_{H}+\chi^2_{\rm fs}\;.
\end{equation}
We use the MCMC method to find the best value of parameters as well as their uncertainties. The results are summarized in Tab.(\ref{tab-res-sm}) and Tab.(\ref{tab-res-cl}) for smooth and clustering cases respectively. In addition, the 1$\sigma$ and 2$\sigma$ confidence regions of the free parameters are presented in Figs.(\ref{fig:res-sm}) and (\ref{fig:res-cl}). 

\begin{table}
	\begin{center}
		\caption{The best value of parameters and the 1-$\sigma$ uncertainties for smooth case.}
		\begin{tabular}{ |l  |c|c|}
			\hline	
			Parameter &  CPL & BA\\
			\hline
			{$\Omega_m       $} & $0.2815\pm 0.0073  $ &$   0.2823\pm 0.0076    $\\
			
			{$h              $} & $0.6965\pm 0.0056   $&$  0.6957\pm 0.0055    $\\
			
			{$w_0            $} & $-0.896\pm 0.079 $&$   -0.908\pm 0.069   $\\
			
			{$w_1            $} & $  -0.50^{+0.41}_{-0.35}  $&$-0.27^{+0.23}_{-0.17}    $\\
			
			{$\sigma_8       $} & $     0.754\pm 0.018       $&$     0.753\pm 0.018     $\\
			\hline
		\end{tabular}
		\label{tab-res-sm}
	\end{center}
\end{table}

\begin{table}
	\begin{center}
		\caption{The best value of parameters and the 1-$\sigma$ uncertainties for clustering case.}
	\begin{tabular}{ |l  |c|c|}
	\hline	
 Parameter &  CPL & BA\\
 \hline
 {$\Omega_m       $} & $0.2814\pm 0.0078          $ &$0.2819\pm 0.0073          $\\
 
 {$h              $} & $0.6964\pm 0.0057          $&$0.6963\pm 0.0053          $\\
 
 {$w_0            $} & $-0.900^{+0.079}_{-0.091}  $&$-0.912\pm 0.065           $\\
 
 {$w_1            $} & $-0.48^{+0.44}_{-0.36}     $&$-0.26^{+0.21}_{-0.18}     $\\
 
 {$\sigma_8       $} & $0.757\pm 0.017            $&$0.757\pm 0.017            $\\
 \hline
\end{tabular}
		\label{tab-res-cl}
	\end{center}
\end{table}
  
  Our results almost show the same confidence regions for the free parameters $\Omega_m$, $h$ and $\sigma_8$. However the DE parameters are constrained slightly different and interestingly, the area of confidence regions for the BA model are smaller than the CPL for both smooth and clustering cases. To quantify this, the DE figure of merit (FoM) is often defined as $\frac{1}{\Delta w_0\Delta w_1}$ where $\Delta w_0$, $\Delta w_1$ are uncertainty of the parameters at 1$\sigma$ level. A large value of FoM means a better constrain and our results indicate that the BA model provide a better constrain than the CPL in both smooth and clustering cases. The FoM for BA parametrization is $\sim$ 70 while it is  $\sim$ 30 for the CPL model (the difference for smooth and clustering cases is very small).  
  
  In order to check the consistency of our models with the observational data,  the corrected Akaike Information Criterion (AIC) has been used. This quantity is given by
  \begin{equation}\label{eq:AIC}
  {\rm AIC}=\chi^2_{min}+2n_{\rm fit}\;,
  \end{equation}
where $n_{\rm fit}$ is number of the free parameters. To compare two models, the pair differences $\Delta$AIC$={\rm AIC}_{y}-{\rm AIC}_{x}$ has to be computed. Two models with $|\Delta$AIC$|\le 2$ are consistent while $|\Delta$AIC$|\ge 6$ indicates a strong evidence against the model with larger AIC. 
In Tab.(\ref{tab:aic}), the AIC values for the two parameterizations as well as the $\Lambda$CDM model are shown. 
 \begin{table}
 	\begin{center}
 		\caption{The AIC value for our models.}
 		\begin{tabular}{ |l  |c|}
 			\hline	
 			Model &  AIC \\
 			\hline
 			{$\rm SCPL$} & $728.9     $ \\
 			\hline
 			{$\rm SBA$} & $728.4          $\\
 			\hline
 			{$\rm CCPL$} & $727.8  $\\
 			\hline
 			{$\rm CBA$} & $727.1    $\\
 			\hline
 			{$\rm \Lambda CDM$} & $728.3    $\\
 			\hline
 		\end{tabular}
 		\label{tab:aic}
 	\end{center}
 \end{table}   
 
According to the AIC criterion, all models are consistent with the data and there is no positive or strong evidence against these models compare to the $\Lambda$CDM. However our results show the DE clustering provide a smaller AIC compare to the smooth cases which indicates the data slightly prefer DE clustering, but of course it's not significant with current data. Notice that a similar conclusion reported in \citep{Basilakos:2014yda,Mehrabi:2014ema,Mehrabi:2015hva} which indicates a good agreement of our results with other works. As we mentioned above, the future observational data, for example based on the \emph{Euclid}, are expected to improve the quality of data significantly and thus the validity of DE clustering will be tested in the near future. 

\section{Conclusion}\label{conclude}
To summarize, we studied the growth of matter perturbations by considering the possibility of DE perturbations. We consider CPL and BA parameterizations with equal number of free parameters but different redshift dependency and integrate the relativistic linear equations to realize the evolution of DM and DE perturbations.
Since from previous works, it was not clear how DE clustering affects the growth rate in different EoS parameterizations, we select the BA model with a different redshift dependency to investigate and compare to the CPL as well as the $\Lambda$CDM.
 Moreover, in contrast to the CPL parametrization, the BA model gives a finite value at far future times ($z\rightarrow-1$). We obtained the relative difference of the Hubble parameter which depends on the free parameters. For instance, for  parameters $w_0=-0.9,w_1=0.2$ the Hubble parameter of the BA (CPL) model is around 6$\%$ (4$\%$) larger than the $\Lambda$CDM case. So for the same values of free parameters the Hubble parameter in these two models differs around $\sim 2\%$ due to the different redshift dependency.
  
We examined both smooth and clustering DE cases within the framework of CPL and BA parameterizations, and calculated the relative difference of growth rate to show the effect of DE clustering on the scenario of structure formation in the Universe. The growth rate is larger (smaller) than the $\Lambda$CDM in the smooth DE cosmology for $w<-1$ ($w>-1$). In contrast to this,  DE clustering gives a larger (smaller) growth index for $w>-1$ ($w<-1$). We observed $(1-2)\%$ difference between the parameterizations analyzed in this work and the $\Lambda$CDM when we fix $w_1=0.2$ and allow $w_0$ varying in range $(-1.3,-0.7)$ for smooth DE ($1-8\%$ for clustered DE). We also examined how the growth rate changes with respect to  $w_1$ parameter and it is found that the difference between our parameterizations and concordance $\Lambda$CDM model is very small in the case of smooth DE. However, in the case of DE clustering, the difference might be as large as $17\%$ ($6\%$) when $w_0=-0.9$ and $w_1=0.6$ for the BA (CPL) model. 
Notice that the difference between the growth rate of these parameterizations is due to the different redshift dependency of the EoS so our results indicate that in the case of clustered DE, understanding the exact functional form of EoS parameter is crucial quantity but its effect is not significant in the case of smooth DE. A combination of weak lensing and the peculiar velocity observations can distinguish between the smooth or clustered DE \citep{2011PhRvD..84h3523S,Asaba:2013mxj}.

In order to check the consistency of our models with observation, we use current available data including  SN Ia (the JLA sample),  Planck CMB, BAO, the Hubble parameter and the growth rate $f\sigma_8$ to put constrain on the cosmological parameters. The MCMC method has been used to obtain the best fit values of the parameters as well as their uncertainties. We obtained almost the same confidence regions for $\Omega_m,h,\sigma_8$ pairs. However, this is not the case for $w_0,w_1$ pair. We measure the FoM of DE as $\sim70$ for the BA and $\sim 30$ for the CPL parameterizations with a tiny difference between the smooth and clustering cases. This means, apart from type of DE (smooth or clustering) scenarios the BA parameterization provides a tighter constrain compare to the CPL. Hence based on this result, we suggest the BA parametrization instead of the CPL as a totally better approximation for the EoS parameter. Finally on the basis of the AIC criterion, these two parameterizations are consistent with observations as equally as the $\Lambda$CDM cosmology. As a comparison between clustering and smooth DE scenarios, we found a smaller AIC for clustering cases which indicates that the observational data slightly prefer clustered DE models. 



 \bibliographystyle{apsrev4-1}
  \bibliography{ref}

\end{document}